\documentclass[%
superscriptaddress,
amsmath,amssymb,
aps,
pra,
floatfix,
onecolumn
]{revtex4-2}

\usepackage{graphicx}
\usepackage{pgfplots}
\pgfplotsset{compat=1.18} 
 
\usepackage{dcolumn}
\usepackage{bm}
\usepackage{hyperref}

\newcommand{\POLIMI}{Dipartimento di Fisica - Politecnico di Milano, piazza Leonardo da Vinci 32, 20133 Milano (Italy)}
\newcommand{\IFNCNR}{Istituto di Fotonica e Nanotecnologie - Consiglio Nazionale delle Ricerche, piazza Leonardo da Vinci 32, 20133 Milano (Italy)}

\begin{document}

\title{Tailoring the birefringence of femtosecond-laser-written \\ multi-scan waveguides in glass}

\author{Roberto Memeo}
\affiliation{\IFNCNR}
\author{Davide Piras}
\affiliation{\POLIMI}
\author{Roberto Osellame}%
\affiliation{\IFNCNR}
\author{Andrea Crespi}
\email{andrea.crespi@polimi.it}
\affiliation{\POLIMI}
\affiliation{\IFNCNR}

\date{\today}

\begin{abstract}
Femtosecond-laser direct waveguide writing is progressively emerging as an alternative to conventional  techniques to develop complex photonic devices, for applications ranging from classical and quantum information processing, to sensing and metrology. Laser written waveguides typically offer low modal birefringence, thus preserving coherence of polarization-encoded information. Integrated waveplates have been reported, as waveguides with tilted birefringence axis, but with limited flexibility in terms of achievable rotation angle, birefringence magnitude or control in the modal shape.
Here we investigate the multi-scan approach to realize low-loss optical waveguides in fused silica substrate with controlled modal birefringence. We show that by tuning the horizontal and vertical shifts between subsequent scans we can independently change both the magnitude and the axis inclination of the birefringence, while keeping efficient mode coupling with standard fibers. 
\end{abstract}

\maketitle

\section{Introduction}

Femtosecond laser direct writing is a powerful technique for rapid prototyping of waveguide devices in glass and crystals, with numerous and diverse applications demonstrated in recent years \cite{tan2021,corrielli2021,piacentini2022,li2022}. 
Single-mode optical waveguides are often realized with this technique as single irradiated tracks, which in general may present complex refractive index profile in the cross-section, supporting only the fundamental guided mode for two orthogonal polarization states. Femtosecond laser written waveguides typically present a low modal birefringence in the order of $10^{-6}-10^{-4}$, which makes them suitable to propagate arbitrary polarization states without loosing coherence \cite{sansoni2010}. Depending on the laser writing approach, one can set the birefringence in the above interval, producing devices with very different properties with respect to light polarization. On the one hand, large interferometric networks indeed can be built that present remarkable polarization insensitivity in their operation \cite{pentangelo2024}. On the other hand, different components have been demonstrated in the recent years, inscribed with femtosecond laser pulses, that also manipulate the polarization state of light by exploiting the modal birefringence. These include polarization beam splitters \cite{fernandes2011pbs,crespi2011integrated}, integrated waveplates \cite{fernandes2011ret,corrielli2014,heilmann2014}, or polarization controlled couplers \cite{corrielli2018,wang2018}.

Tuning the writing parameters may allow a certain control on the waveguide birefringence \cite{fernandes2011ret}. However, to gain a more complete control on the magnitude and axis direction of the waveguide birefringence, either a dedicated beam shaping technique is employed on the writing beam, to alter the symmetry of the writing process \cite{corrielli2014}, or additional tracks are inscribed close to the guiding one \cite{fernandes2012,heilmann2014} to induce well-oriented stress in it. The former method has limitations in terms of the maximum achievable tilt angle, while the latter may generate unwanted spurious effects due to optical modes possibly guided by or around the additional tracks.

Writing waveguides with a single irradiated track has the disadvantage of a non-trivial control of the cross section, because laser-induced modifications are produced by an interplay of different microscopic mechanisms and generally result in articulated shapes.
The waveguide cross-section, on the other hand, may be constructed accurately by juxtaposing several irradiation tracks.

Nasu et al. \cite{nasu2005low} first demonstrated low-loss waveguides in fused silica with a nearly square cross-section, by irradiating several tracks with a horizontal separation of 0.4~$\mu$m. The tight control on the waveguide cross-section allowed them to achieve optimum modal overlap with channel waveguides fabricated in silica by photolithographic techniques, and the laser-written waveguides could be used as low-loss interconnects.
The multi-scan technique can be used advantageously also to obtain multi-mode waveguides with controlled profiles, and to build devices with interconnected single-mode and multi-mode waveguide segments. Interestingly, for instance, Thomson et al. demonstrated a 16-mode photonic lantern \cite{thomson2011}, while Douglass et al. produced an arrayed waveguide grating \cite{douglass2018}. More recently, Sun et al. provided impressive demonstrations of the capabilities of the multi-scan technique in the realization of mode converters and mode rotators, also exploiting waveguides with a three-dimensional twisted geometry \cite{sun2022}. 
Multi-scan waveguides typically yield low polarization-dependent losses \cite{nasu2005low}, hence they are ideal to propagate light in different polarization states. However, their birefringent properties have not been investigated in depth.

In this work, we study how to finely tune the magnitude of the modal birefringence of femtosecond laser written waveguides, realized via the multi-scan technique, and their birefringence-axis direction. We show that by changing the horizontal scan separation it is possible to vary considerably the birefringence magnitude, while the inclination angle of the birefringence axis is controlled easily by the vertical separation between the scans. These features allow us to demonstrate integrated waveplates able to efficiently convert the incoming polarization state.

\section{Methods}

In this study, waveguides are inscribed in 1-mm thick fused-silica substrates (Foctek inc.), by irradiation with a commercial Yb-based femtosecond laser (LightConversion Pharos) at $\lambda$~=~1030~nm wavelength. In particular, the pulse train repetition rate is set at 50~kHz and the laser beam is focused by a 20$\times$ microscope objective (LEICA Achroplan). The waveguide cross-section of the desired shape is produced by stacking 20 laser scans with different separations, as discussed in the next Sections.

For femtosecond-laser-written waveguides, when only the fundamental spatial mode is supported (typically referred to as single-mode waveguides), each guided mode can be seen as the composition of two orthogonally polarized optical modes, which, due to the low birefringence, are rather similar in propagation constant and confinement properties. In addition, due to the low refractive index contrast, it is reasonable to approximate the two optical modes as linearly polarized \cite{snyder1978mow}. Hence, a waveguide behaves as a birefringent medium, producing a phase delay $\delta \phi$ between two orthogonal linearly-polarized modes according:
\begin{equation}
\delta \phi = \frac{2 \pi}{\lambda} (n_s - n_f) L - 2 \pi m
\label{eq:phase}
\end{equation}
where $\lambda$ is the wavelength, $n_s$ and $n_f$ are the effective refractive indices of the slower and faster propagating modes ($n_s \geq n_f$), $L$ is the waveguide length and $m$ is an integer number that brings back $\delta \phi$ in a 2$\pi$-wide phase domain. In the following, we will evaluate the birefringence $b \geq 0$ as:
\begin{equation}
b= n_s - n_f
\end{equation}
We may thus call \textit{fast} axis the polarization direction of the mode with index $n_f$, and define its orientation as the angle $\theta_b$ between this direction and the vertical one (orthogonal to the substrate surface).

The birefringence of the fabricated waveguides is characterized as in Refs.~\cite{sansoni2010,corrielli2014}. Namely, different polarization eigenstates are injected in the waveguide input, and for each different input state we reconstruct the Stokes vector of the output state. In other words, we perform a polarization tomography of the output state for an overcomplete set of inputs. We then fit numerically the axis angle $\theta$ and phase delay $\delta \phi$ that best describe the experimental data.

In our characterization experiments, we use a $\lambda$~=~1550~nm diode laser as a source of coherent light, collimated in free space. After a first passage through a Glan-Thomson polarizer, the polarization state is tuned by means of a sequence of a half- and a quarter-waveplate, properly rotated. Light is injected in the waveguide under study through a 10$\times$ microscope objective. Output light is collected by a 25$\times$ objective and analyzed by a quarter waveplate and a rotatable polarizer, and sent to a power meter or a CMOS camera.

\begin{figure}
\centering
\includegraphics{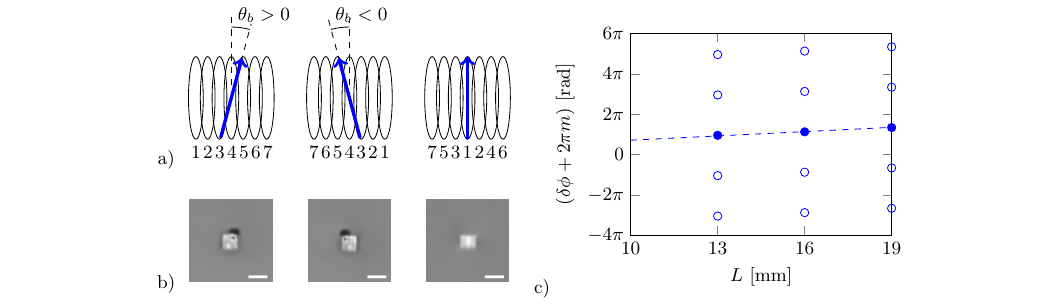}
\caption{\label{fig:stdWg} a) Schematic representation of the effect of the scan writing order on the orientation of the fast axis of the waveguide (depicted as the blue arrow in the drawings). b) Optical microscope images of the corresponding waveguides: the darker dot on the cross section is related to optical damage of the substrate with resulting stress accumulation. Scalebar corresponds to 10~$\mu$m. c)  Cut-back data of the evaluated phase delay between two orthogonal polarization states propagating inside the waveguide for different sample lengths. The plot reports each data point together with its 2$\pi$-periodicity; the dashed line is of the form of Eq.~\ref{eq:phase}, with $\lambda = 1550$~nm and $m=0$. }

\end{figure}

\section{Results}
\subsection{Optimization of standard waveguides}

As a starting point for our study, we optimize the irradiation parameters to obtain a multi-scan waveguide that supports a single spatial mode at $\lambda$~=~1550~nm, with low propagation loss and high mode overlap with standard single-mode fiber (SMF-28e). Taking as an initial reference the work by Nasu et al. \cite{nasu2005low}, we employ 20 scans separated horizontally by 0.4~$\mu$m. Optimal parameters are identified as 190~nJ pulse energy and 1~mm/s translation speed, which are adopted for all the waveguides discussed in this work. Resulting waveguide cross-section is about 8~$\times$~7~$\mu$m$^2$.

Importantly, we find that the writing order of the scans strongly influences the inclination of the birefringence axis. In particular, we tested the three orderings schematized in Fig.~\ref{fig:stdWg}a. Proceeding from left to right or from right to left results in a tilted axis ($\theta_b \simeq \pm 23^\circ$). Instead, a vertical axis within the accuracy of the measurement setup ($< 0.5^\circ$) is obtained with the symmetric inner-to-outer order (the last configuration in Fig.~\ref{fig:stdWg}a), which results in 0.33~dB/cm propagation loss and a mode size of 12.8~$\times$~12.5~$\mu$m$^2$ (comparable for horizontal and vertical polarizations). Estimated coupling loss with a single-mode fiber is 0.13 dB.
A possible explanation of this phenomenon could lie in the fact that the repeated overlap between consecutive laser scans generates stress accumulation and relaxation phenomena over and around the irradiated regions. Hence, the writing order of subsequent scans could determine a preferential direction for the accumulated stress, resulting in a rotation of the birefringence axis. This anisotropy is strongly reduced when writing alternately one scan on the left and one on the right.

As far as the birefringence is concerned, the characterization procedure reported in the previous Section provides without ambiguity only the value of $\delta \phi$, i.e. the phase delay modulo $2\pi$ between the two orthogonal polarization states propagating in the structure. To retrieve the birefringence value $b$ from Eq.~\ref{eq:phase}, one needs to identify the value for $m$. To resolve the ambiguity, we employed a cut-back method by measuring the waveguide-related phase delay while progressively cutting the sample to shorter lengths. For each measurement, we plotted the evaluated phase $\delta \phi$ together with the other possible values differing by multiples of $2\pi$ (i.e. for different values of the integer $m$), as reported in Fig.~\ref{fig:stdWg}c. We then identified the best-fitting straight line through the points, knowing that for $L=0$ the phase should be null. The slope of the fitted line is the birefringence value, which is found to be $b_0 = (5.5 \pm 0.2) \times 10^{-5}$, thus solving the initial ambiguity.
In the following birefringence measurements, we did not perform such a thorough analysis with the cut-back method, but we assumed smooth variations of the birefringence value from a reference $b_0$, obtained for the waveguide written at 0.4~$\mu$m horizontal scan separation.

\begin{figure}[t]
\centering
\includegraphics{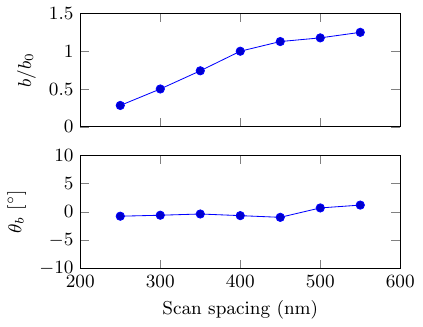}
\caption{\label{fig:scanVar} Evaluated birefringence modulus in units of $b_0$ (birefringence of the reference waveguide of Fig. \ref{fig:stdWg}), and inclination angle $\theta_b$ of the fast axis, as a function of the horizontal scan spacing, for waveguides with a rectangular cross-section.}
\end{figure}

\begin{figure}[t]
\centering
\includegraphics{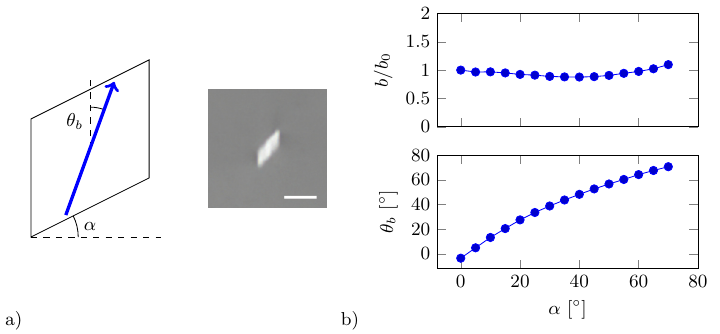}
\caption{\label{fig:tilted} a) Schematic representation and optical microscope picture of a waveguide with tilted cross-section ($\alpha = 35^\circ$). Scalebar corresponds to 10~$\mu$m; $\alpha$ and $\theta_b$ are shown in the drawing with their positive orientation. b) Evaluated birefringence modulus in units of $b_0$, and fast-axis axis $\theta_b$, as a function of the cross-section inclination angle $\alpha$.}
\end{figure}

As a second experiment, we studied how the scan separation influences the birefringence and the other waveguide properties. Starting from the aforementioned reference parameters, different horizontal scan separations were tested, ranging from 0.25~$\mu$m to 0.55~$\mu$m with 0.05~$\mu$m steps. In order to deposit a comparable amount of energy per unit of cross-section area, each waveguide was inscribed by scaling the inscription speed when decreasing the scan separation. In other words, closer scans are characterized by proportionally higher translation speeds and vice versa. The symmetric inner-to-outer writing order was preserved in order to keep the waveguides optical axis vertical. As reported in Fig.~\ref{fig:scanVar}, the birefringence axis $\theta_b$ remains indeed vertical for all the tested separations as expected, whereas its modulus becomes larger with increasing lateral scans displacement.

This trend looks counterintuitive when considering the results of Ref.~\cite{fernandes2012}, where the modal birefringence was observed to increase as the additional tracks written at the sides of the main waveguides were brought closer. Actually, the horizontal displacements we are considering here between scans ($< 1\;\mu$m) are much smaller than those studied in Ref.~\cite{fernandes2012} ($> 10\;\mu$m), and indeed comparable to the physical width of a single track. One possible explanation could be that, in this case, the inscription of a new track partially overlapped to the previous one lowers the local birefringence in the overlap region, thus resulting in a modal birefringence of the overall structure that depends inversely on the amount of overlap. Alternatively, this could be ascribed to a strong influence of the inscription speed of the tracks on the induced birefringence.

The $1/e^2$ mode-field diameters change of less than 20\% while varying the separation of the scans, with the modes turning slightly elliptical as the tracks get closer. In fact, these minimal variations do not affect significantly the coupling efficiency with standard single-mode fibers, with estimated coupling losses remaining lower than 0.2~dB for all the tested configurations (see Supplementary Figure~1).

\subsection{Integrated optical waveplates}

\begin{figure}[t]
\centering
\includegraphics{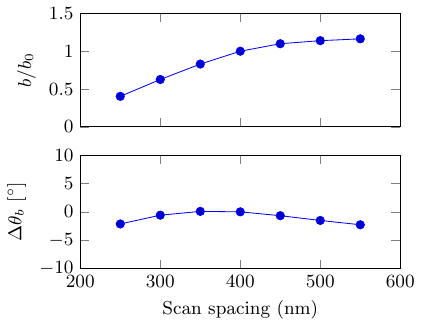}
\caption{\label{fig:tiltedVarScan} Evaluated birefringence modulus in units of  $b_0$, and variation of the inclination of the fast axis, as a function of the horizontal scan spacing, for a waveguide with fixed cross-sectional tilt $\alpha = 35^{\circ}$ (corresponding to $\theta_{b,0} \simeq - 45^{\circ}$).}
\end{figure}

Rotated integrated waveplates can be produced by introducing also a vertical shift between subsequent scans, so that the waveguide cross-section takes the shape of a parallelogram. To investigate how the rotation of the birefringence axis is influenced by this change in the cross-section shape, we fabricated several waveguides consisting in 20 scans of fixed horizontal scan spacing, and different amounts of vertical shift between subsequent scans. In this way, the resulting waveguide cross-sections is characterized by a  tilt angle $\alpha$ with respect to the reference rectangular one (see Fig.~\ref{fig:tilted}a). In particular, values for $\alpha$ were tested ranging from $0^{\circ}$ to $70^{\circ}$ with $5^{\circ}$ steps. 

For each track, we measured the modulus of the birefringence with respect to that of a waveguide with standard rectangular cross section (namely, $b_0$) and its inclination axis $\theta_b$. The results, reported in Fig.~\ref{fig:tilted}b, show a strong correlation between the birefringence axis and the angle $\alpha$, with the value of $\theta_b$ increasing smoothly (in modulus) as the cross-section tilt of the waveguides becomes larger.  The fast axis of the waveguide tends to lean towards the major geometric diagonal of the cross-section (Fig.~\ref{fig:tilted}a illustrates the sign conventions).
Furthermore, the birefringence magnitude only slightly oscillates around the reference value $b_0$. This shows that one can flexibly tune the birefringence axis of the waveguides at will, without significantly affecting its modulus. 

Indeed, we observed experimentally that the birefringence modulus can be changed independently, also in this configuration, by adjusting the horizontal scan separation. In detail, we fixed the cross-sectional tilt $\alpha = 35^{\circ}$ (corresponding to $\theta_{b,0} \simeq - 45^{\circ}$ in the experiment of Fig.~\ref{fig:tilted}), and we varied the horizontal scan separation, ranging from 0.25~$\mu$m to 0.55~$\mu$m. Also in this case, the birefringence magnitude increases smoothly with the scan separation while the axis inclination shows only slight variations from the designed one $\theta_{b,0}$. 
These results demonstrate a good control of the waveguide birefringence properties, decoupling the variations of its modulus to those related to its axis.

As far as the supported modes are concerned, the progressive tilting of the waveguides cross-section results in a change in the mode shape, becoming larger and more elliptical as the angle $\alpha$ increases (see Supplementary Figure 2). Nevertheless, single mode operation is preserved for all tested angles. 
Consistently with the observations reported in the previous Section for rectangular waveguides, when the lateral scan displacement is varied keeping the $\alpha$ angle fixed, the mode size is not altered significantly. Supplementary Figure 3 reports measured $1/e^2$ mode diameters and estimated coupling losses for the waveguides of Fig.~\ref{fig:tiltedVarScan}.

\section{Discussion and Conclusions}

We have presented a novel approach to tune the birefringence properties of femtosecond-laser written waveguides in fused silica glass. Leveraging a lateral multi-scan irradiation technique, the birefringence modulus and its inclination axis can be controlled independently by changing the horizontal separation between subsequent scans and their vertical displacement. 

Differently from the approach of Ref.~\cite{corrielli2014}, here optical waveguides with birefringence axis well beyond $45^{\circ}$ are easily demonstrated. On the other hand, in contrast with the method proposed in Ref.~\cite{heilmann2014}, the presented scheme has the advantage that no additional tracks are needed to tune the waveguide properties. Indeed, here the birefringence is tweaked within the waveguide cross-section, making  our approach particularly practical and versatile when realizing photonic circuits with tightly-packed waveguides, where the presence of additional laser tracks could induce unwanted effects also on nearby irradiated paths. In fact, in our work all the written tracks equally compose the waveguide cross-section in a more gentle overall material modification, where no damaging of the substrate is observed. Furthermore, the manipulation of the birefringence axis inclination is decoupled from the tuning of its modulus, which can be carefully controlled by the horizontal scan separation. 

The presented approach offers unparalleled flexibility for the fabrication of integrated waveplates, useful for the integrated manipulation of arbitrary polarization states of light. These may be applied as polarization rotators in the elaboration of classical signals (e.g. in implementing polarization-diversity schemes), or even to manipulate quantum states of light. Integration with etched microchannels in the same chip, a possibility that is uniquely enabled by the femtosecond laser micromachining technique\cite{piacentini2022},  could open the way to integrated polarization measurements on fluidic samples for biochemical or cellomics applications.

\pagebreak

\begin{acknowledgments}
R.M. and A.C. acknowledge funding from the European Union - NextGenerationEU , under the National Recovery and Resilience Plan (NRPP), mission 4, component 2, investment 1.1, project FEMTO-PRINTER (id. 2022ZZNANK).
\end{acknowledgments}

\bibliography{biblioBire}

\cleardoublepage

\begin{center}

\vfill

\includegraphics{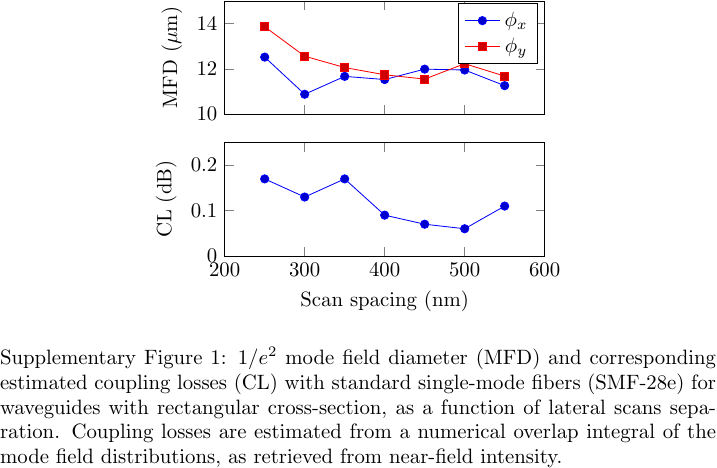}

\vfill

\includegraphics{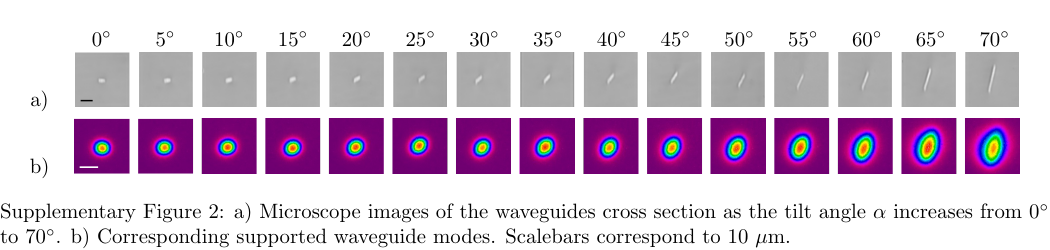}

\vfill

\includegraphics{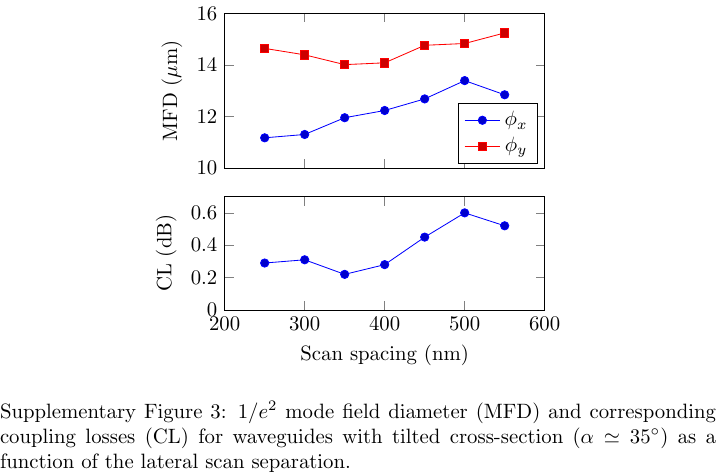}

\vfill
    
\end{center}

\end{document}